\documentclass{article}

\usepackage{arxiv}
\usepackage{float}
\usepackage[utf8]{inputenc} 
\usepackage[T1]{fontenc}    
\usepackage{hyperref}       
\usepackage{url}            
\usepackage{booktabs}       
\usepackage{amsfonts}       
\usepackage{nicefrac}       
\usepackage{microtype}      
\usepackage{graphicx}
\usepackage{natbib}
\usepackage{doi}
\usepackage{tabularx}
\usepackage{array}

\title{From Augmentation to Reconstruction:\\
Guiding the AI Disruption to the Good Place}

 \author{
   \textbf{David M. Rothschild\thanks{All authors are at Microsoft Research.
   Contact: David@ResearchDMR.com} \quad
   Jake M. Hofman \quad
   Markus Mobius \quad
   Brendan Lucier} \\
   \textbf{Eleanor Dillon \quad
   Daniel G. Goldstein \quad
   Nicole Immorlica \quad
   Aleksandrs Slivkins}
 }

\date{Originally submitted May 28, 2026\\
Revised \today}

\renewcommand{\headeright}{Perspective on AI Disruption}
\renewcommand{\undertitle}{Perspective on AI Disruption}
\renewcommand{\shorttitle}{Guiding the AI Disruption to the Good Place}

\hypersetup{
pdftitle={Guiding_AI_Disruption_Good_Place},
pdfsubject={cs.AI, econ.GN},
pdfauthor={David M. Rothschild, Jake M. Hofman, Markus Mobius, Brendan Lucier, Eleanor Dillon, Daniel G. Goldstein, Nicole Immorlica, Aleksandrs Slivkins},
pdfkeywords={AI Disruption, Market Design, Agentic Web, Good Place},
}

\begin{document}
\maketitle
\begin{abstract}
Artificial intelligence feels like it is everywhere, but for most the promised disruption has not yet arrived. The main reason for this is not model capability, nor even the tools built to harness those models, but rather most organizations are using AI to accelerate tasks, inside of workflows that are designed for a pre-AI world. We offer a three-stage lens for AI's impact: Augmentation, Automation, and Reconstruction. The most consequential disruption resides in the third stage, Reconstruction, where workflows and markets are rebuilt for delegation, machine‑to‑machine interaction, continuous monitoring, and auditable constraints. It takes time to reconstruct workflows, as it requires trust and accountability infrastructure, machine-legible and interoperable data and interfaces, and economic incentives that favor reconstruction rather than local optimization. These are the complementary investments that produce the familiar ``productivity J-curve'' of general-purpose technologies. We illustrate this transition through examples in consumer markets, education, news, and coding. Finally, we emphasize a normative point: the agentic future is not predetermined. Leaders must both skate to where the puck is going and actively steer it toward a good place, ensuring innovation delivers welfare gains felt by businesses and consumers around the world.
\end{abstract}

\keywords{AI Disruption \and Market Design \and Agentic Web \and Good Place}

Public discussion of artificial intelligence often overemphasizes current AI performance on a narrow set of familiar tasks. We marvel at models that can ace standardized tests; we love to challenge AI agents' ability to navigate web interfaces and social media. These feats capture attention but they distract from the more consequential impacts to come. The real story is not whether AI can pass exams or navigate our digital environments or surpass benchmarks in our current digital world, but how it will fundamentally reshape how people learn, work, and coordinate their lives. The right questions are whether it helps people acquire new skills, make better decisions, or collaborate more effectively. Today's AI agents that competently browse the web will feel misplaced in a future digital ecosystem that is expressly designed for AI agents to communicate, negotiate, and transact with each other on our behalf.

While it is useful in the short term to explore how AI can augment humans within existing workflows, the greatest disruptions will emerge as we redesign tasks, tools, and environments around AI’s distinctive strengths \citep{bresnahan1996technical}. The transition from asking how well AI augments or automates current tasks within existing workflows to re-imagining what is possible when workflows are rebuilt to leverage AI’s capabilities is central to realizing AI’s long‑term impact \citep{demirer2026chaining}.

History offers a helpful analogy in how early factory electrification replaced steam engines with electric motors, delivering modest productivity gains. Productivity surged only when the factory layouts and production workflows were redesigned to take advantage of electricity’s flexibility and decentralization \citep{david1990dynamo,brynjolfsson2000beyond}. Similarly, today AI is regularly deployed to augment familiar tasks in familiar workflows, such as managing shopping lists, assisting with teaching and school work, and summarizing articles \citep{dillon2025shifting}. These applications can improve quality, reduce costs, and save time, but constrained within familiar workflows they are only capturing a small share of AI's potential. Far larger gains will come when entire workflows are designed from the ground up to optimize for what AI can do both within and between tasks. For example, imagine marketplaces where personal agents continuously coordinate, search, and negotiate on a user’s behalf. Or, education systems organized around persistent, adaptive learning loops that track and build knowledge and skills over time. Or, information ecosystems in which agents monitor sources, evaluate credibility, and synthesize evolving narratives tailored to users’ goals.

\subsection*{Three stages of AI integration}

We start by outlining the three stages of AI integration: Augmentation, Automation, and Reconstruction \citep{rothschild2026agentic}. These stages are not mutually exclusive. In fact, as organizations move tasks and workflows through these stages, they may occupy multiple stages simultaneously. Yet, only the final stage produces genuine structural transformation. These stages map onto the ``productivity J-curve'' of general-purpose technologies \citep{Brynjolfsson2021productivity}, but while that framework describes the aggregate pattern that creates delayed gains, our contribution is to characterize the organizational transitions, from task-level augmentation to workflow redesign, that give rise to it.

\subsubsection*{Stage 1: Augmentation}
In Stage 1, AI enhances discrete tasks currently being performed by humans within existing workflows. This stage dominates AI usage today: drafting emails, summarizing documents, generating code suggestions, answering customer questions, or producing first‑pass analyses. Humans use AI to augment work within tasks, but retain control over sequencing, judgment, and accountability.

Stage 1 spreads quickly because it does not require tasks to be easily connected or workflows to be redesigned. It integrates easily into workflows built for human coordination and oversight. The gains are real, but their overall impact is limited by bottlenecks elsewhere in the workflow. Faster drafting still waits on approvals and faster analysis still feeds slow decisions. Augmentation improves speed and allows users to shift their attention to other tasks, but does not change the workflow structure.

\subsubsection*{Stage 2: Automation}
In Stage 2, routine tasks move ``under the hood'' and require less human awareness. Consider education: a Stage~1 implementation might help a teacher draft a lesson plan or generate practice problems. In Stage~2, these tasks are largely delegated. AI generates lesson plans, creates assessments from that lesson plan, and assists with grading, while students interact with tutors that adapt explanations in real time. Both teachers and students have time freed up to spend on higher-level objectives: structuring learning paths, interpreting results, guiding progress, and targeted learning. For product and operations leaders, Stage 2 is often the first ``real'' productivity step-change: it cuts coordination drag and frees their time for higher-leverage work. Yet, Stage 2 typically is still bogged down by human-era artifacts and workflows: forms, queues, approvals, and meetings. AI works harder inside the system, but the system itself remains largely unchanged.

Stage~2 is best understood as transitional rather than terminal. The key is that this stage is not just about improving the current situation. It also includes substantial complementary investment in data, infrastructure, and organizational processes, much of it intangible and only partially reflected in measured output during the transition. Thus, observed current productivity may understate underlying progress, reflecting early investment that is intangible and imperfectly measured. Further, early gains are often concentrated among initial adopters and may therefore be one-time level shifts rather than sustained improvements in measured output per unit of work. Whether this manifests as a temporary slowdown  (i.e., ``J-curve'') or continued but muted growth (i.e., a ``hockey-stick'' shaped pattern of productivity growth) remains uncertain. This creates a particularly unsettled moment for workers, as investment flows into future workflow reconstruction and their familiar workflows begin to shift under them. During this transition, firms may favor increased investment in capital and partially automated workflows over broad investment in new skills, even as longer-run opportunities may expand for workers.

Many will worry that the disruption ends here with slow or negative growth, and massive worker displacement. The concern is not that automation leads to a ``bad'' outcome per se, but that stopping here risks locking in systems optimized around legacy workflows, limiting the broader gains that come from redesigning how work is organized. As such, Stage 2 is best understood not as a stable endpoint, but as a transitional, suboptimal equilibrium in AI integration.

\subsubsection*{Stage 3: Reconstruction}
Stage 3 is where the long‑anticipated disruption actually resides, and the challenges in reaching it explain why it has not yet emerged at scale. In Reconstruction, workflows are redesigned around AI’s distinctive capabilities: speed, parallelism, memory, continuous monitoring, and machine‑to‑machine interaction. Steps are re‑sequenced, eliminated, or converted into auditable constraints. Governance is engineered directly into workflows rather than layered on after the fact. Human roles shift to align with the higher-level tasks afforded by the automation of routine subtasks. Efficiency gains change the calculus on which forms of work are most valuable, leading to new approaches that were previously considered infeasible. In this process, the activity itself is redefined: what it means to ``shop,'' ``learn,'' or ``code'' changes, rather than simply how those tasks are executed. Critically, the transition depends on institutional conditions: trust, interoperability, workflow redesign, and aligned incentives, that, for most domains, remain underdeveloped.

For AI, this stage of adoption remains early and uneven, with coding among the first domains where elements of reconstruction are emerging. AI-native workflows are beginning to restructure software development around agents, shifting humans upstream into ideation and downstream into curation and editing.

Table~\ref{tab:three_stages_examples} illustrates the progression across four domains. In each case, the shift is not simply from humans doing tasks faster to AI doing the same tasks automatically, but from task support to workflow redesign around delegation, continuous monitoring, and machine-to-machine coordination.

\begingroup
\setlength{\intextsep}{12pt} 

\begin{table}[H]
\centering
\small
\renewcommand{\arraystretch}{1.25}
\setlength{\tabcolsep}{5pt}
\begin{tabularx}{\textwidth}{
  >{\raggedright\arraybackslash}p{2.2cm}
  >{\hsize=.9\hsize\linewidth=\hsize\raggedright\arraybackslash}X
  >{\hsize=.9\hsize\linewidth=\hsize\raggedright\arraybackslash}X
  >{\hsize=1.2\hsize\linewidth=\hsize\raggedright\arraybackslash}X
}
\toprule
\textbf{Domain} & \textbf{Stage 1: Augmentation} & \textbf{Stage 2: Automation} & \textbf{Stage 3: Reconstruction} \\
\midrule

\textbf{Shopping} 
& AI helps draft a shopping list, suggest products, or summarize reviews. The consumer still browses, compares, and decides. 
& Consumers delegate defined tasks (e.g., ``reorder groceries'' or ``find a flight under \$500''), and AI executes the task across platforms with limited oversight. 
& A personal agent continuously manages shopping for the consumer: it anticipates the consumer's needs, compares and coordinates goods and services across many service agents (potentially enabling custom products), negotiates over price and terms, and may commit to a transaction (or delegate back to the consumer). Markets shift from browsing interfaces to discovery and quality assurance necessary for agent-to-agent coordination.\\

\textbf{Education} 
& Teachers may use AI to draft lesson plans or test questions. Students may use AI to help them on their assignments, explain concepts, and generate practice problems for them.
& Teachers may use AI systems that generate assignments from the teacher's lesson plans and manage assessments of the students' work. Students may interact with AI tutors that adapt explanations in real time. 
& Education is reorganized around continuous, personalized learning loops: AI tracks students' achievement, generates tailored lessons for them, assesses their understanding and identify difficulties throughout their lessons, and feeds their results back into curriculum design. The emphasis shifts from static outputs (assignments and tests) to dynamic skill acquisition with continuous evaluation. \\

\textbf{News} 
& AI summarizes articles, highlights key points, and aggregates headlines from multiple sources. 
& Consumers receive personalized daily briefs, in their preferred format, generated from a mix of sources and formats filtered based on their preferences, subscriptions, and past behavior.
& News consumption becomes an interactive, continuously updated dialogue: consumer agents monitor sources, evaluate credibility, cross-check claims, and synthesize evolving narratives tailored to the consumer's goals. The unit of consumption shifts from discrete pieces of content to adaptive, agent-curated understanding.\\

\textbf{Coding} 
& AI provides quick assistance when developers get stuck: suggested code snippets, help debugging errors, and explanations of unfamiliar functions, but humans still write and structure most code.
& AI automates segments of the coding workflow, including code generation, testing, refactoring, and documentation, allowing developers to delegate well-scoped tasks while retaining oversight of any segment of the workflow.
& Software development workflows are restructured around AI agents that can plan, generate, and iterate on code end-to-end; humans shift primarily to problem formulation, architectural guidance, and editing, with agent-to-agent coordination increasingly handling execution. \\

\bottomrule
\end{tabularx}
\setlength{\abovecaptionskip}{4pt} 
\setlength{\belowcaptionskip}{0pt}
\caption{Illustrative progression from augmentation to reconstruction across everyday domains.}
\label{tab:three_stages_examples}
\end{table}

\subsection*{AI Integration in Consumer Markets}

The distinction between stages is vivid in consumer markets, particularly shopping.

Today's digital economy is built for humans to browse, search, and compare offerings through interfaces designed to capture human attention and facilitate human navigation. AI typically appears as a layer on top of these human-optimized systems in the form of recommendations, chatbots, or other personalization tools. These are Stage 1 improvements. More recently, there is increasing interest in autonomous shopping agents that act as personal assistants to which simple shopping tasks can be delegated. These are an initial wave of Stage 2 integration.

A reconstructed, Stage 3 market looks different. In an \emph{agentic market}, consumers delegate tasks to personal assistant agents and firms deploy service agents that represent the pricing, policies, and constraints for their goods and services. The conversion journey shifts from human‑to‑interface interactions to agent‑to‑agent coordination with humans mainly operating at the beginning and end. The consumer expresses goals and constraints; their agent then queries and negotiates with multiple provider agents across the economy. Firms compete less on visibility and more on value, responsiveness, and trustworthiness, communicated between agents in machine‑readable terms. Product recommendation could shift from a centralized service provided by platform intermediaries to an emergent outcome of more decentralized (and open) discovery and quality-assurance systems, enabling more direct, agent-to-agent interaction.

An agentic market is more than a chatbot bolted onto a website. It is a fundamentally different market mechanism, characterized by lower communication friction, faster matching, and new forms of discovery and competition \citep{rothschild2026agentic}.

If agentic markets are increasingly plausible, why do they not yet dominate? While the Stage 3 disruption presupposes Stage 2 integration, the more binding constraints are institutional rather than technical. System-wide change depends on trust and accountability infrastructure necessary for humans to delegate to systems, complementary machine‑legible and interoperable data and interfaces necessary for agents to operate, the design of new agent‑first workflows, and economic incentives that support investment in reconstruction rather than local optimization. These complementary conditions remain underdeveloped in most domains, thus limiting the impact of AI even as agentic capabilities improve. Four constraints to reaching the full disruption stand out.

\paragraph{Trust and accountability remain unresolved at scale.}
Consumers and firms will only delegate meaningful authority to agents if they are both well aligned with their interests and there are clearly assigned and enforced limits on their actions and responsibility. Stakeholders also need auditability and recourse: they must be able to inspect agent behavior, evaluate outcomes of agent activity, respond to failures, and have viable recourse when things go wrong. Until trust infrastructure necessary for these preconditions, which includes instrumentation, monitoring, evaluation, incident response, and liability frameworks, becomes routine, most firms will constrain agents to assistive or advisory roles rather than granting true autonomy.

\paragraph{Data and interfaces lag behind agentic needs.}
Reliable agentic execution requires clean, interoperable data, along with explicit, machine‑readable governance. Many firms' data flows are not cleanly legible for machines to use reliably in autonomous operations, let alone in interactions with external agents. Reliance on human-centric interfaces and brittle data pipelines raises error rates, compounding risks of delegation. Some risk can be offset by stronger harnesses and control layers, but further innovation is necessary before multi‑agent systems can operate robustly at scale.

\paragraph{Workflows remain fundamentally human‑centric.}
Most enterprise systems assume humans mediate interactions, so they are optimized for humans as they, for example, read screens and fill out forms. Agent‑first systems instead assume agents execute routine work, thus they optimize for agentic performance with: APIs, protocols, and machine‑readable rules that allow agents to transact, coordinate, and escalate efficiently without constant human intervention. This is a structural redesign of workflows and infrastructure, not an incremental layer added on top of existing tools.

\paragraph{Incentives favor optimization over reinvention.}
It is safer and easier to use AI to optimize within existing workflows and business models than to rebuild organizations and markets from the ground up. It is difficult for firms to simultaneously invest in transformation while continuing to meet short-term revenue targets, discouraging large-scale reconstruction for long-term gain. Yet the largest gains from AI and agentic systems come from this difficult and costly investment. This creates a misalignment of incentives and capacity: incumbents can afford the upfront costs of new infrastructure but lack incentives to disrupt themselves, while smaller or newer firms have stronger incentives to be transformative but more limited capacity to invest.

This creates a natural question: if Stage~2 delivers real gains at lower risk, why would any firm voluntarily undertake the cost of reconstruction? The answer is that Stage~2 equilibrium is stable only in the absence of competitive pressure. At some point there will be an entrant unburdened by legacy systems or a consortium that solves the coordination problem that demonstrates a reconstructed workflow, rapidly shifting the equilibrium. Moreover, prolonged Stage~2 optimization accumulates organizational debt around legacy architectures, making eventual reconstruction more expensive. The critical question is not whether reconstruction will occur, but whether incumbents will lead it or be disrupted by it.

\subsection*{Guiding the AI Disruption to the Good Place}
In technology, leaders often invoke the hockey metaphor: \emph{skate to where the puck is going}. AI makes that advice feel urgent. But in an agentic economy, skating alone is not enough.

AI will reshape how work is done and how markets function. How large the gains are, whether they are broadly distributed, whether ecosystems remain open, and whether consumers and businesses retain meaningful agency all depend on choices made now \citep{rothschild2026agentic}.

The AI disruption, left to default conditions, risks reinforcing walled gardens where agents are embedded in proprietary ecosystems that preserve friction, lock in users, and concentrate surplus. The large platforms that have the capital and technical capacity to invest in agentic systems are naturally disincentivized from directing the disruption in a way that threatens existing business models. Thus, the default trajectory of this disruption is to harden today’s market structures with more automated versions of the status quo, rather than enable the advanced agentic coordination necessary for broader societal gains. Avoiding this outcome requires deliberate early experimentation with more open, auditable, and transformative designs before lock‑in becomes irreversible.

This is the difference between reacting to where the puck appears headed and \emph{actively directing the puck towards a good outcome}.

A practical way for organizations to prepare for Stage 3 is to invest in underlying infrastructure for advanced agentic coordination, prioritizing long-term impact over short-term features or demos. In short, this means shifting interest and investment from isolated tools to system-level redesign, making deliberate long-range choices about what types of agents operate in the system, how authority is delegated, and how value is created and shared. Or more actionably:
\begin{itemize}
\item Invest in machine-readable data and interfaces that enable agent-to-agent interaction
\item Redesign workflows to be AI-native, rather than layering automation onto existing processes
\item Define clear delegation boundaries and accountability structures for agent-driven decisions
\end{itemize}

We are still early in this transition not just because the technology is still rapidly evolving, but because reconstruction is difficult and costly. Leaders still have time to make explicit choices about what type of economy they wish to build. The central question is not whether AI will produce a meaningful disruption, but whether the disruption will make the economy more open and competitive producing broad-based welfare gains, or hard harden today's power structure into faster and more automated version of the status quo.

AI’s advance is inevitable. Its impact on society is not.

\bibliographystyle{unsrtnat}
\bibliography{sample}

\end{document}